\documentclass[twocolumn]{aastex631}
\usepackage{graphicx}
\usepackage{float}

\usepackage{tablefootnote}

\DeclareUnicodeCharacter{2212}{\ensuremath{-}}

\newcommand{\lsim }{{\lower0.8ex\hbox{$\buildrel <\over\sim$}}}
\newcommand{\gsim }{{\lower0.8ex\hbox{$\buildrel >\over\sim$}}}
\newcommand{\Msun}{\ifmmode {M_{\odot}}\else${M_{\odot}}$\fi}
\newcommand{\Lsun}{\ifmmode {L_{\odot}}\else${L_{\odot}}$\fi}
\newcommand{\Rsun}{\ifmmode {R_{\odot}}\else${R_{\odot}}$\fi}

\shorttitle{A radio companion to UltraCompCAT}
\shortauthors{Dage et al.}

\begin{document}

\title{Radio Continuum Studies of Ultra-Compact and Short Orbital Period X-Ray Binaries }

\correspondingauthor{Kristen Dage}
\email{kristen.dage@curtin.edu.au}

\author[0000-0002-8532-4025]{Kristen C. Dage}
\affiliation{International Centre for Radio Astronomy Research -- Curtin University, GPO Box U1987, Perth, WA 6845, Australia}

\author[0000-0001-8424-2848]{Teresa Panurach}
\affiliation{Center for Materials Research, Department of Physics, Norfolk State University, Norfolk VA 23504, USA }
 
\author[0000-0003-1814-8620]{Kwangmin Oh}
\affiliation{Center for Data Intensive and Time Domain Astronomy, Department of Physics and Astronomy, Michigan State University, East Lansing, MI 48824, USA \\} 
\author[0000-0003-0440-7978]{Malu Sudha}
\affiliation{Department of Physics \& Astronomy, Wayne State University, 666 W. Hancock St, Detroit, MI 48201, USA}

\author[0000-0002-4344-7334]{Montserrat Armas Padilla}
\affiliation{ Instituto de Astrofísica de Canarias (IAC), Vía Láctea s/n, La Laguna 38205, S/C de Tenerife, Spain}
\affiliation{Departamento de Astrofísica, Universidad de La Laguna, La Laguna E-38205, S/C de Tenerife, Spain}

\author[0000-0001-9261-1738]{Arash Bahramian}
\affiliation{International Centre for Radio Astronomy Research -- Curtin University, GPO Box U1987, Perth, WA 6845, Australia}

\author[0000-0002-8294-9281]{Edward M. Cackett}
\affiliation{Department of Physics \& Astronomy, Wayne State University, 666 W. Hancock St, Detroit, MI 48201, USA}

\author[0000-0002-2801-766X]{Timothy J. Galvin}
\affiliation{CSIRO Space \& Astronomy, PO Box 1130, Bentley WA 6102, Australia}

\author[0000-0003-3944-6109]{Craig O. Heinke}
\affiliation{Physics Department, CCIS 4-183, University of Alberta, Edmonton, AB, T6G 2E1, Canada}

\author[0000-0002-8961-939X]{Renee Ludlam}
\affiliation{Department of Physics \& Astronomy, Wayne State University, 666 W. Hancock St, Detroit, MI 48201, USA}

\author[0000-0003-0851-7082]{Angiraben D. Mahida}
\affiliation{International Centre for Radio Astronomy Research -- Curtin University, GPO Box U1987, Perth, WA 6845, Australia}
\author[0000-0002-7092-0326]{Richard M. Plotkin}
\affiliation{Department of Physics, University of Nevada, Reno, NV 89557, USA}
\affiliation{Nevada Center for Astrophysics, University of Nevada, Las Vegas, NV 89154, USA}

\author[0000-0002-7930-2276]{Thomas D. Russell}
\affiliation{INAF, Istituto di Astrofisica Spaziale e Fisica Cosmica, Via U. La Malfa 153, I-90146 Palermo, Italy}

\author[0000-0001-9261-1738]{Susmita Sett}
\affiliation{International Centre for Radio Astronomy Research -- Curtin University, GPO Box U1987, Perth, WA 6845, Australia}

\author[0000-0002-5319-6620]{Payaswini Saikia}
\affiliation{Center for Astro, Particle and Planetary Physics, New York University Abu Dhabi, PO Box 129188, Abu Dhabi, UAE}

\author[0000-0002-8808-520X]{Aarran W. Shaw}
\affiliation{ Department of Physics \& Astronomy, Butler University, 4600 Sunset Ave, Indianapolis, IN 46208, USA}

\author[0000-0003-3906-4354]{Alexandra J. Tetarenko}
\affiliation{Department of Physics and Astronomy, University of Lethbridge, Lethbridge, Alberta, T1K 3M4, Canada}

\begin{abstract}
We present the radio continuum counterparts to the enigmatic ultra-compact X-ray binaries (UCXBs); a black hole or neutron star accreting from a hydrogen-deficient white dwarf donor star, with short orbital periods ($<$ 80 minutes). For the sample of UCXBs hosted by globular clusters (GCs), we search for whether certain GC properties are more likely to enhance UCXB formation. We determine that GCs which host UCXBs are drawn from a distinct population in terms of cluster concentration, core radius and half-light radius, but are similar to other well-studied GCs in metallicity and cluster mass. In particular, UCXB-hosting GCs tend to be on average more compact, with a higher concentration than other GCs, with significantly higher encounter rates. We investigate whether a correlation exists between radio luminosity and orbital period, using new and archival observations. We determine that there is not a clear connection between the two observable quantities.
\end{abstract}


\section{Introduction}  
Ultra-compact X-ray binaries (UCXBs) are black holes (BHs) or neutron stars (NSs) accreting from hydrogen-deficient donor stars \citep{1986ApJ...304..231N}. They are typically defined as having orbital periods less than 80 minutes \citep{Nelemans10a,Heinke13}, although theory suggests that at later stages in their evolution, UCXBs may have orbital periods of up to two hours \citep{vanHaaften12, vanHaaften12b}. They are valuable laboratories to study accretion and binary evolution \citep[e.g.][]{2017MNRAS.467.3556B}, given the availability of rich long-term datasets \citep[][and references therein]{2023A&A...677A.186A}. 

UCXBs can form either through standard common envelope evolutionary channels \citep{Nelemans10a} or, in the more dynamic environment, through physical collisions between NSs and red giants \citep{Ivanova05}.  After formation, gravitational wave emission  plays a major role in their evolution \citep{Chen21,Suvorov21}. UCXBs will also be some of the loudest gravitational wave sirens for the Laser Interferometer Space Antenna (LISA), which has a planned launch in the 2030s \citep{2023LRR....26....2A}, and make up a significant number of the Milky Way's known population of low mass X-ray binaries \citep[LMXBs;]{2024A&A...684A.124F}.

Although unified by the definition of having a short-orbital period and a hydrogen-deficient donor star, the sample of known UCXBs is diverse, including BH candidates (BHC), NSs and accreting millisecond pulsars (AMXPs), both persistent and transient, with some exhibiting X-ray bursts. UCXBs are found both in the field and in globular clusters  \citep[GCs;][and references therein]{2023A&A...677A.186A}. 

\begin{table*}
\label{table:obs}
\caption{Log of ATCA observations.}
\begin{tabular}{llllll}
Name      & R.A         & Dec.         & Obs. Date              & Integration Len (min)   & Secondary \\ \hline
1M 1716-315    & 17:18:47.02 & -32:10:14.54 & 2024-02-13             & 207.7          &  1714-336  \\
1RXS J170854.4-321857 & 17:08:54.27 & -32:19:57.13 & 2024-03-12             & 238.9          & 1714-336  \\
1RXS J171824.2-402934  & 17:18:24.14 & -40:29:33.04 & 2024-03-07             & 219.6           & 1714-397  \\
1RXS J180408.9-342058  & 18:04:08.37 & -34:20:51.19 & 2024-03-11             & 222.9           & 1741-312  \\
4U 1626-67    & 16:32:16.79 & -67:27:39.3  & 2024-02-23,2024-02-24  & 237.1           & 1619-680  \\
AX J1538.3-5541   & 15:38:14    & -55:42:13.6  & 2024-04-03             & 137.5           & 1511-55   \\
IGR J17062-6143 & 17:06:16.3  & -61:42:40.5  & 2024-02-12, 2024-02-13 & 222.7          &  1658-62   \\
IGR J17254-3257  & 17:25:24.8  & -32:57:15    & 2024-02-28,2024-02-29  & 198.1         & 1714-336  \\
IGR J17494-3030  & 17:49:23.62 & -30:29:59    & 2024-03-04             & 192.8           & 1741-312  \\
SAX J1712.6-3739   & 17:12:36.77 & -37:38:41.0  & 2024-03-12             & 222.5         & 1714-397  \\
SAX J1806.5-2215 & 18:06:32.17 & -22:14:17.32 & 2024-03-10             & 235       & 1752-225  \\
SLX 1737-282   & 17:40:42.83 & -28:18:08.4  & 2024-03-20             & 225.8          & 1741-312  \\
SLX 1744-299   & 17:47:25.89 & -30:00:01.6  & 2024-03-09             & 222.4          &  1741-312  \\
XMMU J174716.1-281048  & 17:47:16.16 & -28:10:48    & 2024-03-11             & 238.5          & 1748-253  \\
XTE J1807-294   & 18:06:59.8  & -29:24:30    & 2024-03-02,2024-03-03  & 208.2          & 1817-254  \\
XTE J1751-305   & 17:51:13.49 & -30:37:23.4  & 2024-02-15             & 126.2          &  1741-312 
\end{tabular}
\end{table*}
UCXBs are typically discovered and studied in X-rays. However, radio continuum observations have played a key role in identifying UCXB 47 Tuc X-9 as a BHC \citep{Miller-Jones15}, although over 90\% of UCXBs and UCXB candidates are NSs \citep[][and references therein]{2023A&A...677A.186A}. 

Both accreting BHs and NSs can emit synchrotron emission (associated with relativistic jets) that can be probed by radio continuum observations. These jets are thought to be correlated with the X-ray emission (associated with the accretion disk or electron corona), in both NSs and BHs, with the former occupying a different region of the X-ray/radio luminosity plane compared to BHs \citep{Merloni03, Falcke04, Gallo18, Panurach21}. However, for NSs, these correlations are difficult to interpret as properties of the NS like the magnetic field strength, may impact the jets, and therefore what is observed in radio \citep[e.g.][]{Migliari11, Tudor17, Gusinskaia17, vandenEijnden21}. %Therefore, radio studies are an important component to untangling jet formation in unique systems like UCXBs, and provide a valuable metric to compare them to other BH and NS systems.
  For many of these systems which are transient, the jet is launched in outburst, and radio observations of the quiescent continuum can provide a useful baseline for future observations. We also note that in several systems, a jet is detected in quiescence \citep[e.g.][]{2008MNRAS.389.1697C,2019MNRAS.488..191G,2020MNRAS.493L.132T,2021MNRAS.503.3784P}. 

%Uniquely, as UCXBs have white dwarf donors, one might expect the accretion disk to be helium, with higher abundances of carbon/oxygen than other XRBs (Juett 2001), and this different composition may possibly carry through to the jets/outflows observed in radio (van den eijnden 2021).

In this paper, we provide a radio companion to \cite{2023A&A...677A.186A}'s UltraCompCAT; combining new observations, literature studies of UCXBs and searching all sky radio surveys. We explore whether the orbital period can be correlated with X-ray and radio luminosity, and the observed properties of globular clusters that host UCXBs.

\section{Data and Analysis}
UltraCompCAT lists 49 UCXB, UCXB candidate and short orbital period LMXBs, 24 of which have orbital periods reported \citep{2023A&A...677A.186A}. Candidate UCXBs, selected as such via their unique accretion signatures \citep[e.g.][]{2007A&A...465..953I}, are also included. We obtained 16 new observations from the Australia Telescope Compact Array (ATCA; \citealt{Wilson11}), compile 16 radio observations from the literature of UCXBs with orbital period measurements, and search archival radio surveys of a further 9 sources. For the sake of consistency, in all further analysis, we adopt the peak 2-10 keV X-ray luminosity reported by \cite{2023A&A...677A.186A}. 
%All of the literature UCXBs report X-ray and radio luminosities that were taken near-simultaneously, except for ?0916 and 0513, and in these cases, we adopt the X-ray luminosities based on assumptions made in Tetarenko 2018. 

We note that many of these systems are transient \citep[e.g.,][]{2024arXiv240718867H}; radio observations taken during quiescence can serve as upper limits and provide useful reference points for future detections when the source enters outburst again. Additionally, for systems where the nature of the compact object remains unclassified, a deep radio detection could help constrain its properties and assess whether it is a BHC. %of the systems not in outburst will be useful to act as quiescence upper limits, should the sources be detected again. In the case of the unclassified source, a deep radio detection would provide constraints on the suitability as a BHC. %even a radio non-detection may provide clarification  radio luminosity to the two UCXB BH candidates \citep{Miller-Jones15,2021MNRAS.507..330S} even with a radio non detection. %For the persistent sources, where one might expect radio emission to be produced, we treat the radio observations as upper limits. %We caveat that unlike the X-ray/radio measurements from the literature literature, the radio measurements of these sources will not be simultaneous with the X-ray.
\subsection{New Observations}
We observed 16 UCXBs with ATCA, using a sparse sampled 6km array configuration with the 4cm Compact Array Broadband Backend (CABB) receiver \citep{Wilson11}. CABB can independently observe two intermediary frequencies, each with 2048 channels with 1MHz widths, in this case at 5.5 GHz and 9.0 GHz (Table \ref{table:obs}). The bandpass and absolute flux calibrator  for all  observations was PKS 1934-638. %The time varying gains and leakage calibrator is indicated in Table \ref{obs} for each source.

We used standard analysis reduction techniques with CASA\footnote{\url{https://casaguides.nrao.edu/index.php/CASA_Guides:ATCA_Advanced_Continuum_Polarization_Tutorial_NGC612-CASA4.7}} ver 5.8.0-109 \citep{CASA2022} to calibrate and image the 5.5 GHz and 9.0 GHz bands separately, using the CASA task \textsc{tclean}. 
We selected a Brigg’s robustness parameter of 1 to preference image sensitivity over spatial resolution for these data. Additionally, data were imaged and cleaned with a second order Taylor expansion (nterms=2) to account for the large fractional bandwidth and source spectral variance.  We performed both bandpass and absolute flux calibration for all observations using ATCAs preferred calibrator PKS 1934-638 \citep{2016ApJ...821...61P}. Time varying gains and on-axis leakage corrections were derived by regular observations of bright and compact sources near our object targets. We list these calibrators in Table \ref{table:obs}.  None of the target UCXBs sources were detected in these images, and we stacked both frequencies to make a deeper image at 7.25 GHz. We used CARTA \citep{2021zndo...4905459C} to calculate the RMS at their position to report 3$\sigma$ upper limits, which are reported in Table 2.

For all sources, we computed radio luminosity by assuming a flat spectrum, $L_R= 4\pi \nu S_\nu  d^2$, where $S_\nu$ is the flux density, $\nu$ is the frequency, and $d$ is the distance to the source reported by \cite{2023A&A...677A.186A}. For distances estimated using thermonuclear bursts, we used those derived under the assumption
of helium-fueled bursts.  In cases where multiple distances were reported, we took the He distance measurement method for the sake of consistency, as more sources in the catalog had distance measured via that method, although we note this may produce a systematic uncertainty in the luminosity.

\begin{table*}
\label{table:atca_uplims}
\caption{RMS upper limits measurement from 7.25 GHz stacked images for quiescent sources. Sources marked with $\dagger$ have a bright source nearby affecting the noise. $\alpha$ was detected in outburst by \cite{Gusinskaia17}, source $\beta$ was detected by \cite{Iacolina10}. }
\begin{tabular}{lllll}
Name                            & P/T & Class       & RMS (Jy) & $<$ Flux (erg/s/cm$^2$)\\ \hline % &$<$  $L_R$ (erg/s) 
1M 1716-315                          & T   & NS        & 8.6$\times10^{-6}$      & 1.9$\times10^{-18}$  \\%&  1.07$\times10^{28}$      \\
1RXS J170854.4-321857                       & P   & NS        & 7.2$\times10^{-6}$       & 1.6$\times10^{-18}$\\%  & 6.10$\times10^{27}$      \\
1RXS J171824.2-402934                        & P   & NS        & 8.2$\times10^{-6}$       & 1.8$\times10^{-18}$ \\% & 7.97$\times10^{27}$      \\
1RXS J180408.9-342058  $\dagger$ $^{\alpha}$ & T   & NS        & 4.5$\times10^{-5}$       & 9.7$\times10^{-18}$ \\% & 3.91$\times10^{28}$      \\
4U 1626-67                          & P   & NS        & 1.8$\times10^{-5}$       & 3.9$\times10^{-18}$ \\% & 3.02$\times10^{28}$      \\
	AX J1538.3-5541  $\dagger$               & T   & ?         & 1.9$\times10^{-5}$       & 4.1$\times10^{-18}$ \\% & 3.13$\times10^{28}$      \\
IGR J17062-6143                       & T   & NS (AMXP) & 1.5$\times10^{-5}$       & 3.3$\times10^{-18}$\\%  & 2.12$\times10^{28}$      \\
IGR J17254-3257                        & P   & NS        & 1.1$\times10^{-5}$      & 2.3$\times10^{-18}$  \\%& 5.86$\times10^{28}$      \\
IGR J17494-3030                        & T   & NS (AMXP) & 1.0$\times10^{-5}$       & 2.3$\times10^{-18}$  \\%& 1.73$\times10^{28}$      \\
SAX J1712.6-3739 $\dagger$               & P   & NS        & 6.9$\times10^{-5}$       & 1.5$\times10^{-17}$  \\%& 4.14$\times10^{28}$      \\
SAX J1806.5-2215                       & T   & NS        & 1.1$\times10^{-5}$       & 2.3$\times10^{-18}$ \\% & 1.05$\times10^{28}$      \\
SLX 1737-282                         & T   & NS (AMXP) & 1.0$\times10^{-5}$       & 2.2$\times10^{-18}$ \\% & 6.85$\times10^{27}$      \\
SLX 1744-299 $\dagger$               & P   & NS        & 2.8$\times10^{-4}$       & 6.1$\times10^{-17}$ \\% & 3.78$\times10^{29}$      \\
XMMU J174716.1-281048$\dagger$               & T   & NS        & 2.6$\times10^{-4}$       & 5.6$\times10^{-17}$ \\% & 4.75$\times10^{29}$      \\
XTE J1751-305  $^{\beta}$             & T   & NS        & 1.1$\times10^{-5}$       & 2.4$\times10^{-18}$ \\% & 2.05$\times10^{28}$      \\
XTE J1807-294                         & T   & NS (AMXP) & 6.0$\times10^{-6}$       & 1.3$\times10^{-18}$  \\%& 1.00$\times10^{28}$     
\end{tabular}
\end{table*}

\subsection{Literature Observations}

Information about the UCXBs in UltraCompCAT with previous radio observations and reported orbital periods are reported in Table 3. 
\begin{table*}
\label{literature}
\caption{Properties of UCXBs with X-ray luminosity, radio luminosity and orbital period. In the case of literature sources, we report both the X-ray and radio luminosity, and convert the X-ray luminosity to the 2-10 keV regime using \textsc{pimms} (https://cxc.harvard.edu/toolkit/pimms.jsp), assuminfg a spectrum with a power-law index of 1.7. For any new sources, which are persistent, we report the peak X-ray luminosity as compiled by \cite{2023A&A...677A.186A}. Orbital periods  and distances are taken from the literature compilation by \cite{2023A&A...677A.186A}, except for the distance to 1E 1603.6+2600, which we adopt from \cite{2004ApJ...608L.101H}. All of these sources are either persistent, or the measurements were taken in outburst (denoted with P/T). The last four sources are short-orbital period LMXBs. Sources marked with $\dagger$ are located in a GC. }
\begin{tabular}{llllllll}
Name               & Class  &P/T & d (kpc)  & P$_{orb}$ (min) & $L_R$ (erg/s)     & $L_X$ (erg/s) & Ref.     \\ \hline
4U 1728-34         & NS     &P &4.4  & 10.87   (?)        & 6.8$\times 10^{28}$          & 6.8$\times 10^{35}$      & \cite{diaztrigo2017}            \\
4U 1820-303   $\dagger$       & NS       &P &8.0& 11.42           & 8.8$\times 10^{28}$          & 3.9$\times 10^{37}$      & \cite{2021MNRAS.508L...6R}                \\
4U 0513-40    $\dagger$       & NS       &P & 12.0&17              & $<$5.5$\times 10^{28}$        & 2.1$\times 10^{36}$      &  \cite{Machinetal1990}       \\
2S 0918-549        & NS     &P   &3.9& 17.4            & $<$5.2$\times 10^{28}$       & 2.0$\times 10^{35}$      & \cite{vandenEijnden21}      \\
4U 1543−624        & NS      &P&9.2  & 18.2            & $<$7.2$\times 10^{27}$        & 7.4$\times 10^{36}$      & \cite{ludlam19}           \\
4U 1850-087    $\dagger$      & NS    &P    &7.4& 20.6            & 4.6$\times 10^{28}$          & 6.8$\times 10^{35}$      & \cite{Lehtoetal1990} \\
M15 X-2     $\dagger$         & NS      &P &11.0 & 22.58        & 1.1$\times 10^{29}$      & 9.2$\times 10^{35}$      & \cite{2011ATel.3378....1M}  \\
47 Tuc X-9    $\dagger$         & BH? &P    &4.5  & 28.18           & 3.4$\times 10^{27}$          & 3.6$\times 10^{33}$      &    \cite{Miller-Jones15}     \\
IGR J17062-6143          & NS (AMXP)&P &7.3& 37.97         & $<$2.1$\times 10^{28}$ & 3.5$\times 10^{35}$      & This work                   \\
4U 1626 -67            & NS &P     &8  & 41.54          & $<$3.0$\times 10^{28}$       & 2.3$\times 10^{36}$      & This work                   \\
XTE J1751-305            & NS (AMXP)&T &       8.5&42.42  & $<$1.1$\times 10^{28}$       & 2.3$\times 10^{32}$      &  \cite{Iacolina10} \\
XTE J0929-314            & NS (AMXP)&T & 10&43.58         & 1.4$\times 10^{29}$          & 3.2$\times 10^{36}$      & \cite{2002IAUC.7893....2R}                \\
IGR J16597-3704     $\dagger$ & NS (AMXP) &T&7.2& 45.97           & 1.5$\times 10^{28}$          & 1.8$\times 10^{36}$      & \cite{Tetarenko18}             \\
4U 1916-053        & NS      &P  &7.6& 49.75           & $<$1.8$\times 10^{29}$        & 3.9$\times 10^{36}$      & \cite{1986ApJ...310..172G}  \\
4U 0614+091        & NS      &P &2.6& 51.3            & 1.7$\times 10^{28}$          & 1.4$\times 10^{36}$      & \cite{Migliari10}        \\
Swift J1756.9-2508 & NS       &T & 8&54.70         & $<$2.5$\times 10^{28}$        & 1.2$\times 10^{37}$      & \cite{Tudose09}                 \\
HETE J1900.1–2455  & NS (AMXP) &T&3.8& 83.25         & 1.1$\times 10^{30}$          & 6.8$\times 10^{35}$      &\cite{2005ATel..530....1R}         \\
1E 1603.6+2600     & NS &P      &$\sim$ 7.5 & 111.04          & $<$7.3$\times 10^{28}$       & 2.8$\times 10^{34}$      & This work                  \\
IGR J17379–3747    & NS (AMXP)&T & 8.5&111.65          & 2.7$\times 10^{29}  $        & 1.5$\times 10^{36}$      &  \cite{vandenEijnden18}       \\
4U 1812-12         & NS       &P &3.0& 114(?)             & $<$2.9$\times 10^{28}$       & 4.1$\times 10^{35}$      & This work                  
\end{tabular}
\end{table*}
%Van den Eijnden 2021 performed a radio census of neutron star X-ray binaries, including four UCXBs. These four were not detected in radio.  

\subsection{Archival Surveys}
We searched for UCXBs without prior radio information in UltraCompCAT using the Rapid ASKAP
Continuum Survey (RACS) and the Very Large Array Sky Survey (VLASS). RACS is an all sky survey with the 
Australian SKA Pathfinder Telescope which 
 began in April 2019 \citep{2007PASA...24..174J,2021PASA...38....9H}. The current data release has a median noise of 165 $\mu$Jy $\rm bm^{-1}$  \citep{2020PASA...37...48M,2023PASA...40...34D}.
 
VLASS is an all sky survey carried out on the Karl G. Jansky Very Large Array \citep{2011ApJ...739L...1P}, beginning in September 2017, with a noise level of 0.07mJy  $\rm bm^{-1}$  \citep{2020PASP..132c5001L}. We searched both RACS and VLASS for potential counterparts to the UCXBs in Table 4. None of the UCXBs were detected. %except for HETE J1900.1–2455, which was detected in all bands, however we note that \cite{2005ATel..530....1R} identified a low frequency background source, and we prefer to use their radio measurements of the source in outburst in the following analysis.

We also searched the Milky Way ATCA VLA Exploration of Radio Sources in Clusters (MAVERIC) survey catalogs of 38 Galactic GCs \citep{Shishkovsky20, Tudor22}. As discussed in \cite{Paduano21}, NGC 6652 B was observed twice by the VLA.  4U 1850-087 in NGC 6712 was observed in 1989 by \cite{Lehtoetal1990}, and was detected again by the VLA in 2014 \citep{Shishkovsky20}. % with a 5GHz flux density of 88.4 $\pm 2.3\micro$Jy. This corresponds to a radio luminosity of 2.88 $ \pm 0.75\times 10^{27}$ erg/s, placing it roughly an order of magnitude lower than the measurement by \cite{Lehtoetal1990}.  
Radio and X-ray variability of this source is discussed in greater detail in \cite{Panurach21,2023ApJ...946...88P}

\begin{table*}
\label{table:racs}
\caption{Radio non-detected sources marked with * from VLASS (2-4 GHz), all other sources from RACS-Low (888 MHz).}
\begin{tabular}{llllll}
Name     & P/T & Type & RMS (Jy)     & $<$ Flux (erg/s/cm$^2$)  \\ \hline %$<$ $L_R$ (erg/s)
1E 1603.6+2600 $^*$              & P   & NS   & 1.2$\times 10^{-4}$ & 1.1$\times 10^{-17}$     \\    %       & 5.65$\times 10^{27}$      \\
MAXI J1957+032 $^*$                & T   & NS   & 1.8$\times 10^{-4}$ & 1.6$\times 10^{-17}$        \\    %    & 4.74$\times 10^{28}$      \\
4U 1812-12                       & P   & NS   & 1.0$\times 10^{-3}$ & 2.7$\times 10^{-17}$   \\    %         & 2.87$\times 10^{28}$      \\
AX J1754.2-2754                  & T   & NS   & 1.1$\times 10^{-3}$ & 2.9$\times 10^{-17}$       \\    %     & 2.88$\times 10^{29}$      \\
1RXH J173523.7-354013             & P   & NS   & 9.0$\times 10^{-4}$ & 2.4$\times 10^{-17}$        \\    %    & 2.58$\times 10^{29}$      \\
SLX 1735-269                     & P   & NS   & 1.1$\times 10^{-3}$ & 2.9$\times 10^{-17}$       \\    %     & 3.13$\times 10^{29}$      \\
XMMU J181227.8-181234 $\dagger$    & T   & NS   & 3.8$\times 10^{-3}$ & 1.0$\times 10^{-16}$   \\    %     & 8.88$\times 10^{30}$      \\
RX J1709.5-2639                   & T   & NS   & 2.7$\times 10^{-3}$ & 7.2$\times 10^{-17}$       \\ %    & 7.26$\times 10^{29}$      \\
1A 1744-361                       & T   & NS   & 1.4$\times 10^{-4}$ & 3.7$\times 10^{-18}$      %      & 3.67$\times 10^{28}$     
\end{tabular}

\end{table*}
\section{Results and Discussion} \label{sec:results}

%\begin{figure}
 %   \centering
  %  \includegraphics[width=0.5\textwidth]{Lr_porb.pdf}
   % \caption{ Radio luminosity versus orbital period. Each system is color coded by accretor (NS, AMXP, BH candidate).Only transient UCXBs with X-ray and radio taken simultaneously during outburst, or persistently accreting sources are displayed here.}
    %\label{fig:lrporb}
%\end{figure}

%\begin{figure}
 %   \centering
  %  \includegraphics[width=0.5\textwidth]{LxLr_porb.pdf}
 %   \caption{ X-ray versus radio luminosity for all UCXBs. Only transient UCXBs with X-ray and radio taken simultaneously during outburst, or persistently accreting sources are displayed here.}
  %  \label{fig:lrlx}
%\end{figure}
\begin{figure*}
\begin{tabular}{ll}
\includegraphics[scale=0.5]{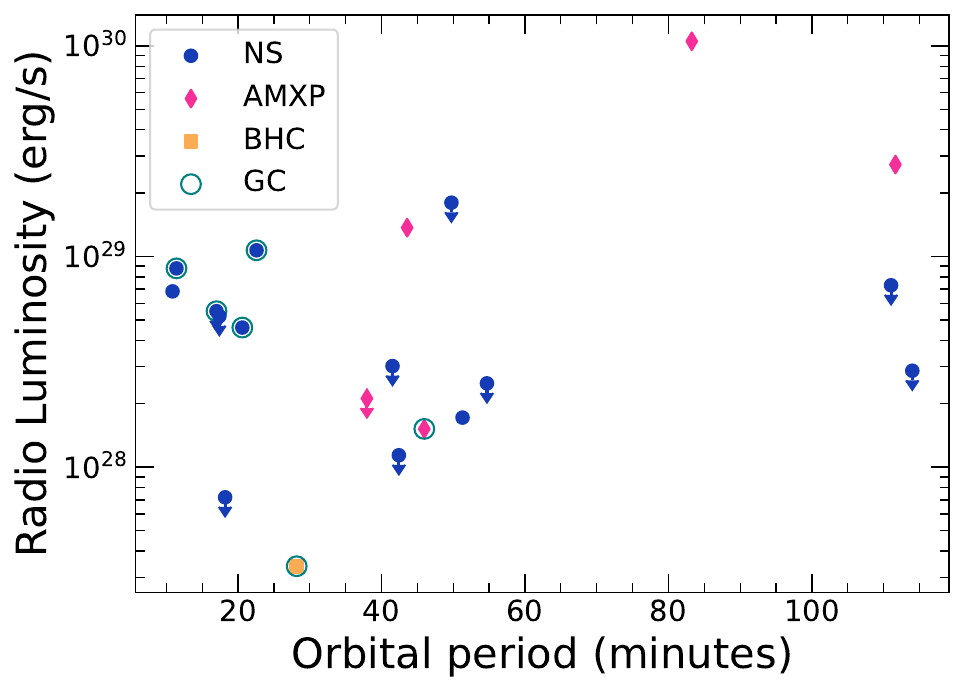}
&
\includegraphics[scale=0.515]{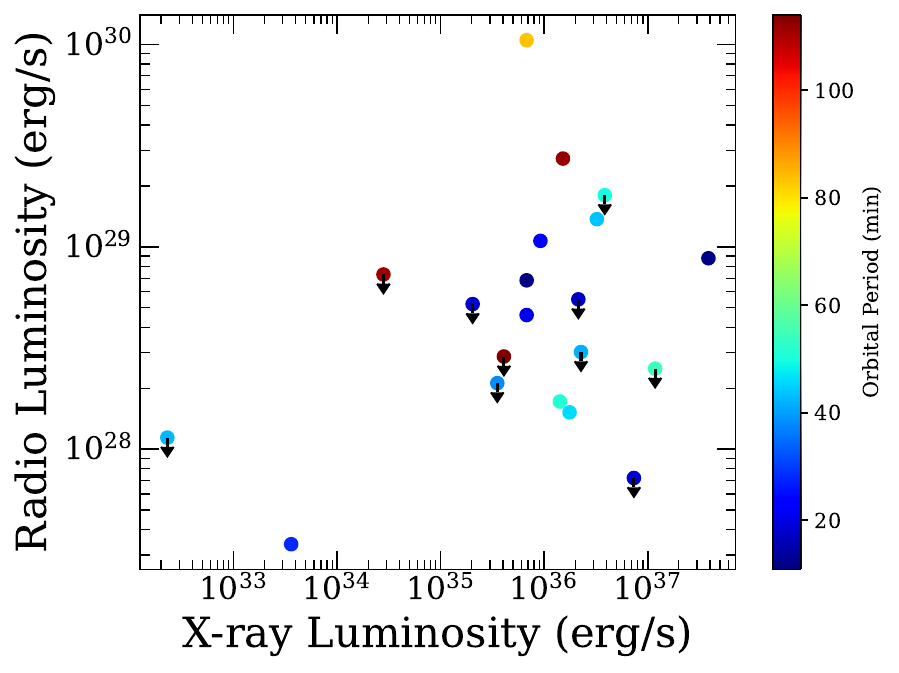}
\end{tabular}
\caption{Only transient UCXBs with X-ray and radio taken simultaneously during outburst, or persistently accreting sources are displayed here. Left: Radio luminosity versus orbital period. Each system is color coded by accretor (NS, AMXP, BHC), and sources located in a GC are circled. Right: X-ray versus radio luminosity for all UCXBs. We do not see a strong correlation between radio, X-ray and orbital period for this sample. }
\label{Fig:32_321}
\end{figure*}
\subsection{Black Hole Candidates}
Four systems in the UltraCompCAT sample are not confirmed as NS accretors (i.e. they do not
exhibit pulsations or thermonuclear X-ray bursts): 47 Tuc X-9, IGR J17285-2922, Swift J0840.7-3516, and AX J1538.3-5541. Of these, radio observations and detection of a radio counterpart have been used by \citealt{Miller-Jones15} (47 Tuc X-9) and \citealt{2021MNRAS.507..330S} (IGR J17285-2922) to suggest a BH accretor. A similar study by \cite{2021A&A...650A..69C}, where no radio emission was detected, is unable to discriminate between a BH and NS in the case of Swift J0840.7-3516. 

\begin{table*}
\label{table:clusterpvalues}
\caption{Cluster property vs p-value from Anderson-Darling test.}
\centering
\begin{tabular}{lllll}
Cluster Property  & AD p-value & Mean (UCXB hosts) & Mean (all other GCs) & Ref. \\ \hline
%Age               &      0.67    &11.86 &12.1  \\
log(Mass)         &      0.63     &5.31& 5.37 & \cite{2018MNRAS.478.1520B}\\
Metallicity       &      0.38    &-1.15& -1.30 & \cite{2018MNRAS.478.1520B} \\
Concentration     & 0.02       &1.97 &1.61 &\cite{Harris96,harris10} \\
Core Radius       & 0.03   &0.28& 0.53  &\cite{Harris96,harris10}   \\
Half-Light Radius &  0.02 &1.27 &1.66 &\cite{Harris96,harris10} \\
log($\Gamma$) &0.001&2.71 &1.95 & \cite{2013ApJ...766..136B}
\end{tabular}
\end{table*}
For AX J1538.3-5541, we did not detect a radio counterpart above $3.13\times10^{28}$ erg/s. Like \cite{2021A&A...650A..69C}, our upper limit is not able to distinguish a BH from a NS. 

% \begin{figure*}
%     \centering
%     \includegraphics[width=0.95\linewidth]{newsrc_lrlx_plot_bhcand.jpg}
 %    \caption{Lr/Lx for BH candidates.}
  %   \label{fig:enter-label}
 %\end{figure*}
%\subsection{Actively Accreting Neutron Stars}
%We group the persistently accreting neutron stars and examine them as a population. 

 %\begin{figure*}
  %   \centering
   %  \includegraphics[width=0.95\linewidth]{newsrc_lrlx_plot_ns.jpg}
   %  \caption{Lr/Lx for NS}
   %  \label{fig:enter-label}
 %\end{figure*}
\subsection{Quiescent Neutron Stars}
Many of the UCXBs in our sample are transient sources, where the radio observations did not take place in outburst. Many of these systems are AMXPs. Here, we report our RMS measurements as quiescent upper limits for future comparison, should they go into outburst again (Tables 2,4).
%\subsection{Accreting Millisecond X-ray Pulsars}
% I know nothing about AMXPs

 %\begin{figure*}
    % \centering
   %  \includegraphics[width=0.95\linewidth]{newsrc_lrlx_plot_amxp.jpg}
   %  \caption{Lr/Lx for AMXPs}
   %  \label{fig:enter-label}
 %\end{figure*}

\subsection{Period versus radio luminosity}
 \cite{Tetarenko18} searched for a correlation between orbital period and radio/X-ray luminosity from 12 UCXBs to assess whether the orbital period plays a role in jet regulation (and hence radio luminosity). However, due to the small and diverse sample, where the systems were likely in different evolutionary stages, they were unable to find any correlation between radio luminosity and orbital period. We update this study with new detections or upper limits of 20 UCXBs (10 detected). For the upper limits (10 sources), we only include persistent UCXBs where one might expect to detect radio emission, and exclude the transient UCXBs where radio observations were not taken in outburst.

\begin{figure*}
    \centering
    \includegraphics[width=0.85\linewidth]{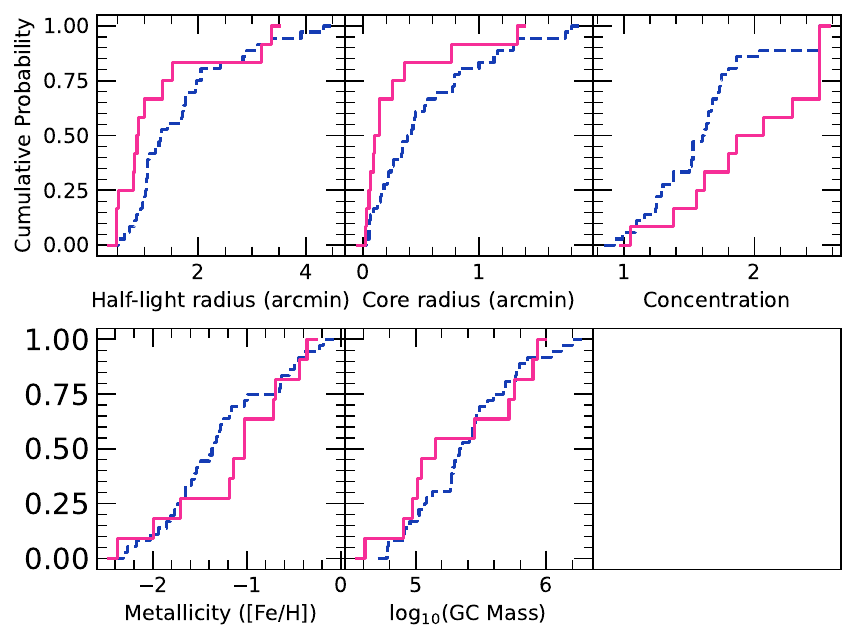}
    \caption{Cumulative distribution function of GC properties (structural parameters, metallicity and cluster mass) with (pink solid line) and without (blue dashed line) detected UCXBs. }
    \label{fig:gc_cdf}
\end{figure*}

\begin{figure}
    \centering
    \includegraphics[width=3.5in]{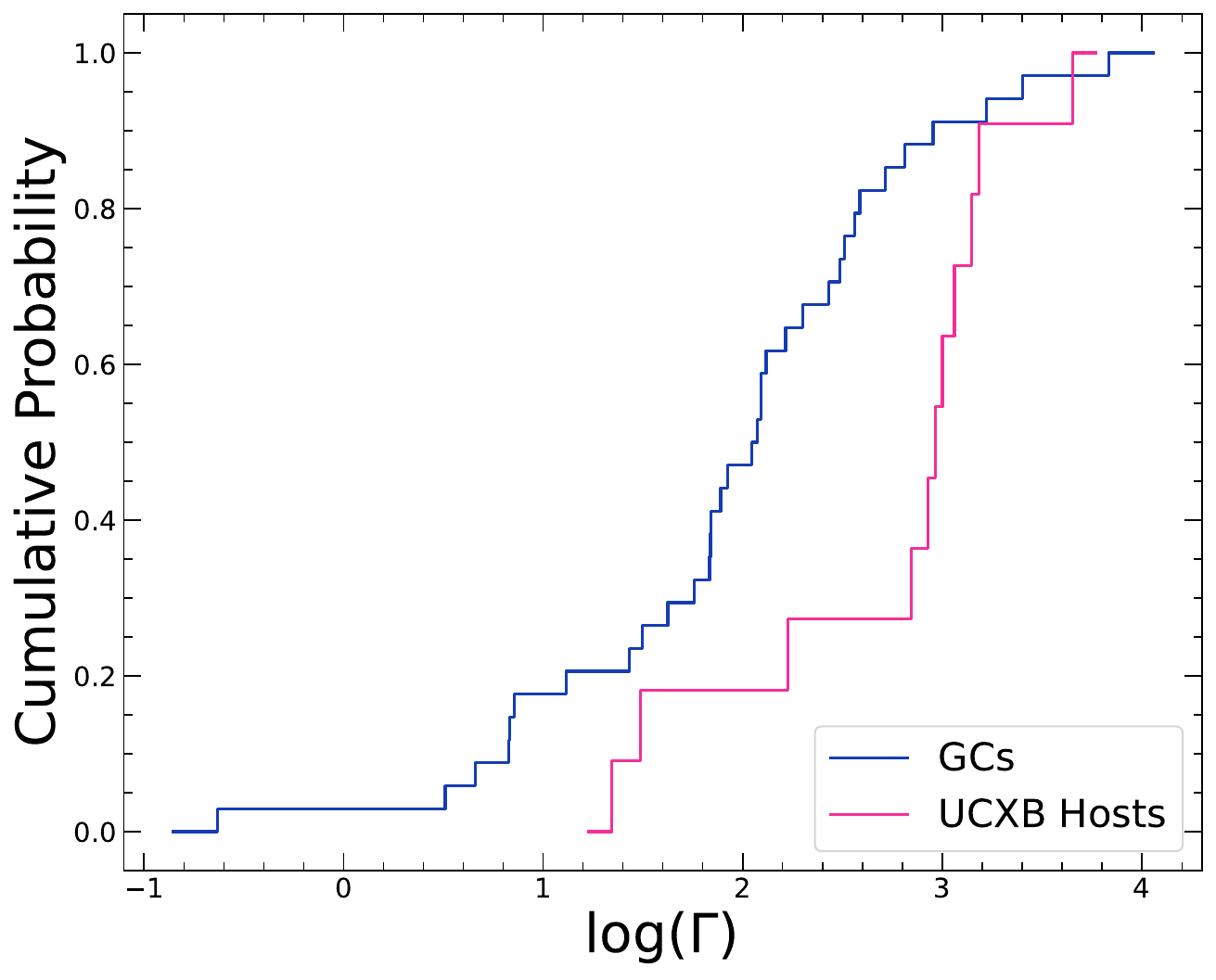}
    \caption{Encounter rate for UCXB-hosting clusters compared to non-UCXB (MAVERIC) GCs. The UCXB-hosting clusters had significantly higher encounter rates than other (non-UCXB) clusters. }
    \label{fig:cdfgamma}
\end{figure}
 Similar to \cite{Tetarenko18}, we search for evidence of a correlation between radio luminosity and orbital period, and find no strong evidence (Spearman rank correlation statistic of 0.14 and p-value 0.55 for the entire sample, statistic of 0.31 and p-value of 0.38 for detections only). We display X-ray versus radio luminosity, color-coded by orbital period in Figure \ref{Fig:32_321} (right panel).
The left panel of Figure \ref{Fig:32_321} shows the orbital period versus radio luminosity.  
\subsection{Globular Cluster vs Field UCXBs} We investigate whether specific properties of GCs (such as metallicity,
mass, and structural characteristics), play a role in the formation and evolution of UCXBs. To do so,
we compare the 11 globular clusters that host UCXBs with a sample of GC that do not contain any
known UCXBs.
UCXBs have been found in 11 globular clusters. Of these, GC dynamics are known to influence the formation of stellar exotica  \citep{2022ApJ...934L...1K,2024MNRAS.532..259O,2025arXiv250106037W}. We compare the observable GC properties from \cite{Harris96,harris10} and  \cite{2018MNRAS.478.1520B}, along with encounter rates from \cite{2013ApJ...766..136B}. We restrict the ``non-UCXB'' comparison sample to 36 clusters from the MAVERIC survey \citep{2020ApJ...901...57B}, as we can consider these to be a complete set of GCs without known UCXBs (more poorly studied GCs may contain as-of-yet undiscovered UCXBs.)%, but the X-ray and radio coverage of the MAVERIC survey is such that similar UCXBs to our sample should have been discovered if present in these clusters)
We analyzed properties of the 11 clusters hosting UCXBs to determine if there are certain cluster parameters (metallicity, mass, cluster structure) that may enhance or influence UCXB evolution. We adopt the calculated encounter rates from \cite{2013ApJ...766..136B}. Of the 11 UCXB-hosting GCs, five are core collapse globular clusters. These literature values are presented in Table 6.
\begin{table*}
\caption{Observed GC properties of the 11 UCXB hosting GCs. Core radius (r$_c$), half-light radius (r$_h$) and concentration values are from \cite{Harris96, harris10}, metallicity and GC mass from \cite{2018MNRAS.478.1520B}, and encounter rates are from \cite{2013ApJ...766..136B}. Core collapsed GCs are denoted with (c).}
\label{table:gcs}
\begin{tabular}{lllllll}
Cluster Name & Concentration & r$_c$ (arcmin) & r$_h$ (arcmin) & Metallicity ({[}Fe/H{]}) & log$_{10}$ GC Mass & Encounter Rate \\
NGC 104      & 2.07          & 0.36                 & 3.17                       & -0.72                        & 5.93              & 1000           \\
NGC 1851     & 1.86          & 0.09                 & 0.51                       & -1.18                        & 5.45              & 1530           \\
NGC 2808     & 1.56          & 0.25                 & 0.8                        & -1.14                        & 5.89              & 923            \\
NGC 6256 (c) & 2.5           & 0.02                 & 0.86                       & -1.02                        & 5.05              & 169            \\
NGC 6293 (c) & 2.5           & 0.05                 & 0.89                       & -1.99                        & 5.15              & 847            \\
NGC 6440     & 1.62          & 0.14                 & 0.48                       & -0.36                        & 5.75              & 1400           \\
NGC 6624 (c) & 2.5           & 0.06                 & 0.82                       & -0.44                        & 5.01              & 1150           \\
NGC 6652     & 1.8           & 0.1                  & 0.48                       & -0.81                        & 4.61              & 511            \\
NGC 6712     & 1.05          & 0.76                 & 1.33                       & -1.02                        & 4.98              & 24.2           \\
NGC 7078 (c) & 2.29          & 0.14                 & 1                          & -2.37                        & 5.71              & 3520           \\
Terzan 2 (c) & 2.5           & 0.03                 & 1.53                       & -0.69                        & 4.91              & 22.1          
\end{tabular}
\end{table*}
 Figure \ref{fig:gc_cdf} shows the cumulative distribution function of UCXB hosts versus the other Milky Way GCs for half-light radius, core radius, concentration, metallicity and cluster mass, and report the mean value of each sample. In Table \ref{table:clusterpvalues}, we show the p-values from Anderson-Darling tests comparing the GC properties\footnote{For the null hypothesis that the observed properties of UCXB-hosting GCs and other GCs in the Galaxy come from a common population.}.  We note that although the entire population of Milky Way GCs is relatively homogeneous (e.g. compared to extragalactic globular cluster populations), this comparison suggests that UCXB hosts are drawn from a distinct population in terms of half-light radius, core radius and concentration, but are not distinguished from other GCs in terms of metallicity or mass. We find that the strongest difference is in encounter rates ($\Gamma$), where the UCXB-hosting GCs have much higher encounter rates than the rest of the sample (Figure \ref{fig:cdfgamma}). This potential difference suggests that certain types of GCs may enhance UCXB formation, which can be tested in the future by \textsc{nbody} cluster simulations. 

\section{Summary and Conclusions} \label{sec:summary}
UCXBs are a very special class of LMXBs.  While radio observations are an important tool in understanding the nature of the accretors and accretion physics in these complex systems, we were unable to find a correlation between orbital period and radio luminosity. Thus, at first glance, the orbital period does not seem to be a factor in jet launching. %We also note that many of our non-detections come from quiescent persistent systems, which is the same for LMXBs.

Although this was the largest sample possible to look for correlation between orbital properties and radio luminosity, we acknowledge that NSs are complicated, and other facets, such as inclination angle, and X-ray spectral state, are more likely to affect what we observe in the radio \citep[e.g.][among others]{Tetarenko18,Panurach21}. Next generation facilities like the ngVLA or SKA will have the sensitivity to probe the radio nature of UCXBs to an even deeper scale. 

We explored the properties of GCs which host UCXBs to those which are not known to host UCXBs. We found that UCXB-hosting clusters appear to be drawn from a different population in terms of their concentration (with a slightly higher average concentration than other clusters), core radius (a smaller core radius) and half-light radius  (slightly smaller half-light radius), but were not demonstrably different in their mass or metallicity. We found that currently, only clusters with high stellar encounter rates host UCXBs. This suggests that there may be certain aspects of the host cluster which are more likely to enhance stable UCXB formation. However, given that UCXBs are most often discovered by their transient events, and that BHs are incredibly difficult to observe/classify, we do not know if our knowledge of GC UCXBs is complete, and we caution that this study was drawn from a relatively small population. Probing theoretical formation and evolution pathways of UCXBs, particularly in GCs, would provide valuable insights for whether these cluster properties do indeed influence UCXB growth. 

 \begin{acknowledgments}
The authors thank the anonymous referee for their thoughtful commends, and acknowledge Adelle Goodwin's help with observations.  KCD acknowledges support for this work provided by NASA through the NASA Hubble Fellowship grant HST-HF2-51528 awarded by the Space Telescope Science Institute, which is operated by the Association of Universities for Research in Astronomy, Inc., for NASA, under contract NAS5–26555.  MAP acknowledges support through the Ramón y Cajal grant RYC2022-035388-I, funded by MCIU/AEI/10.13039/501100011033 and FSE+.  AJT acknowledges the support of the Natural Sciences and Engineering Research Council of Canada (NSERC; funding reference number RGPIN--2024--04458).

 The Australia Telescope Compact Array is part of the Australia Telescope National Facility (grid.421683.a) which is funded by the Australian Government for operation as a National Facility managed by CSIRO. We acknowledge the Gomeroi people as the traditional owners of the Observatory site. 
 
This research has made use of the CIRADA cutout service at URL cutouts.cirada.ca, operated by the Canadian Initiative for Radio Astronomy Data Analysis (CIRADA). CIRADA is funded by a grant from the Canada Foundation for Innovation 2017 Innovation Fund (Project 35999), as well as by the Provinces of Ontario, British Columbia, Alberta, Manitoba and Qu\'ebec, in collaboration with the National Research Council of Canada, the US National Radio Astronomy Observatory and Australia’s Commonwealth Scientific and Industrial Research Organisation.
 \end{acknowledgments}
\vspace{5mm}
\facilities{ATCA, ASKAP, VLA}

\software{ CASA \citep{CASA2022}, matplotlib \citep{Hunter07}, NumPy \citep{harris20}, pandas \citep{Mckinney10}, CARTA \citep{2021zndo...4905459C}, R\citep{CRAN} }

\bibliography{references}{}
\bibliographystyle{aasjournal}

\end{document}